\def\etal{{\it et al.~}} 

 \documentclass[preprint]{aastex}
\usepackage{psfig}

\slugcomment{Submitted to the Astronomical Journal}

\shorttitle{Kunkel \etal}
\shortauthors{Magellanic Cloud Periphery Carbon Stars IV. The SMC}

\begin{document}


\title{Magellanic Cloud Periphery Carbon Stars IV:  The SMC}


\author{William E. Kunkel}
\affil{Carnegie Institution of Washington, La Serena, Chile}
\email{kunkel@jeito.lco.cl}
\author{Serge Demers}
\affil{D\'epartement de Physique Universit\'e de Montr\'eal, Montreal, Qc
H3C 3J7, Canada}
\email{demers@astro.umontreal.ca}

\and

\author{M. J. Irwin}
\affil{Institute of Astronomy, Cambridge CB3 0HA, England}
\email{mike@mail.ast.cam.ac.uk}



\begin{abstract}
The kinematics of 150 carbon stars observed at moderate dispersion on the
periphery of the Small Magellanic Cloud are compared with the motions of
neutral hydrogen and early type stars in the Inter-Cloud region.  The
distribution of radial velocities implies a configuration of these stars as
a sheet inclined at 73$\pm$4 degrees to the plane of the sky.  The near side,
to the South, is dominated by a stellar component;  to the North, the far
side contains fewer carbon stars, and is dominated by the neutral gas.  The
upper velocity envelope of the stars is closely the same as that of the gas.
This configuration is shown to be consistent with the known extension of the
SMC along the line of sight, and is attributed to a tidally induced
disruption of the SMC that originated in a close encounter with the LMC some
0.3 to 0.4 Gyr ago.  The dearth of gas on the near side of the sheet is
attributed to ablation processes akin to those inferred by Weiner \& Williams
(1996) to collisional excitation of the leading edges of Magellanic Stream
clouds.  Comparison with kinematic data of Hardy, Suntzeff \& Azzopardi (1989),
Maurice \etal (1989), and Mathewson \etal (1986, 1988) leaves little doubt that 
forces other than gravity play a role in the dynamics of the H I.

\end{abstract}




\section{INTRODUCTION}

The discovery of the Magellanic Stream by Wannier \& Wrixon (1972) suggested
to many that an interaction between the Magellanic Clouds had occurred in the
relatively recent past, and the more detailed observations of the Stream by
Mathewson and his co-workers (Mathewson, Schwarz \& Murray 1977, Mathewson
\& Ford 1984) revealed a structure as though a turbulent wake had broken the
debris into six distinct fragments, extending over 110$^\circ$ on the sky.
Attempts to account for the observations led to both purely tidal scenarios
(Fujimoto \& Sofue 1976, 1977, Davies \& Wright 1977, Kunkel 1979, Murai \&
Fujimoto 1980) as well as others, involving ram pressure (Fujimoto \& Sofue
1976, 1977, Lin \& Lynden-Bell 1977, Mathewson \etal 1987). In these scenarios
the SMC is seen to be the source of the dispersed gas.  Early theoretical
investigations using N-body simulations employed fixed potentials and no gas
dynamics, and met with limited success. No simulation proved able to deal with
the absence of a stellar component in the resulting debris stream (Moore \&
Davis 1994).  By 1976 interpretations involving stars appeared.  Kunkel \&
Demers (1976), and independently Lynden-Bell (1976) noted that the alignment of
the Stream with the distribution of dwarf spheroidal satellites of the Milky
Way followed a great circle on the sky within some ten degrees, leading to the
inference that a close encounter of the Magellanic Clouds with the Galaxy had
stripped debris fragments in the current form of dwarf spheroidal galaxies from
the Clouds well after their initial formation, in the manner described by Alar
and Yuri Toomre (1972).  Kunkel (1979) suggested that the difficulty in
creating a trailing tail without a leading bridge, and further, to explain the
absence of stars in the stream, required an early disruption, perhaps 6 Gyr 
ago, before star formation had advanced significantly, well prior to the 
current encounter between the Clouds and the Galaxy.  Such an event would then 
create gaseous configurations loosely bound to the Clouds until the present.  
Within a year the difficulty presented by the ``tail-only'' scenario was 
elegantly resolved by Murai \& Fujimoto (1980) who proposed that the binary 
character of the Clouds had created a debris fragment torn from the SMC 
perhaps 1.5 Gyr ago, during an encounter with the LMC, which then remained 
trapped in their joint potential well until the recent encounter with the 
Milky Way deposited the fragment into the tail trailing both Clouds, while the 
bridge locus, containing no corresponding fragment, remained unoccupied.  The 
next observational insight came from am investigation of cepheid variables in 
the SMC by Mathewson, Ford, \& Visvanathan (1986, 1988). Their work showed 
that, unlike the LMC, the depth along the line of sight of the SMC Cepheids 
exceeded the SMC tidal diameter, extending to tens of kpc.  Although this 
interpretation was later challenged (Welch {\it \etal} 1987), evidence for an 
unusual extent of the SMC in the line of sight became apparent in photometry 
of the red giant clump (Hatzidimitriou \& Hawkins 1989, Mateo and 
Hatzidimitriou 1992).  In their photometry the extent of the SMC was seen to 
vary from a maximum of 23 kpc toward the NE (ESO/SERC field 52) to as little 
as 10 kpc in the SW (ESO/SERC field 28). More recently Hatzidimitriou, Cannon, 
\& Hawkins (1993) found a gradient in radial velocity associated with the 
vertical spread in the magnitudes of the red giant clump some three degrees 
East of the SMC, thereby confirming the expansion of this portion of the SMC, 
and a start time corresponding to the epoch earlier workers had associated 
with the tidal encounter between the Clouds. They confirmed that this 
extension of stars in the line of sight ranges over more than 15 kpc, and an 
expansion with a velocity gradient of 7 km s$^{-1}$ per kpc.

A search for stellar components of the inter-cloud region (ICR) began with a
survey by Westerlund \& Glaspey (1971) who found a young stellar population 
some 5$^\circ$ to the East of the SMC photocenter (at {$\it l,b$} = 
302.8$^\circ$, --44.3$^\circ$). Some ten years later Kunkel (1980) 
demonstrated the existence of a young population of stars associated with the 
denser ensembles of neutral hydrogen 8$^\circ$ to the East of the SMC. In the 
following decade a complete Schmidt survey revealed that the entire ICR 
extending all the way to the LMC was covered with a thin population of young 
stars of surprisingly uniform age (Demers \& Irwin 1991). All the blue stars 
of the ICR were found to be unexpectedly young, much younger than the time 
scale attributed to the interaction event between the two Clouds (Grondin 
\etal 1992). The presence of an old stellar component in the ICR, older than 
its dynamical time scale, was found at about this time, in the form of carbon 
stars that likewise extended over the entire ICR (Demers, Irwin \& Kunkel 
1993, hereafter Paper I). Some years before the survey of the periphery carbon 
stars was completed, an early assessment of the ICR data then available 
suggested that the ratio of neutral gas to carbon stars, examined as a 
function of radial velocity (and hence distance by virtue of the expansion 
gradient found by Hatzidimitriou {\it \etal} (1993)) declines with decreasing 
radial velocity. Consequently carbon stars were found to dominate at the lower 
velocities, corresponding to the bridge, while H I dominates the tail which 
shows primarily the more positive velocities (Kunkel, Demers \& Irwin 1995).  
This interpretation is the first indication that non-gravitational forces may 
play a non-negligible role in the ICR dynamics.

The study reported here completes the survey of carbon stars over the entire
periphery of the SMC, and augments the interpretive portion of our analysis
with the extensive kinematic observations of Hatzidimitriou {\it \etal} (1997).
Section 2 describes our observational material and the procedures employed to
produce it, Section 3 describes briefly the material incorporated from other
sources, while section 4 presents the data in a variety of geometric 
projections selected to facilitate interpretations.  Section 5 provides a 
description of an interpretive scenario based on self-gravitating N-particle 
simulations modeled on a disruptive interaction with the LMC, from which 
interpretive constraints are inferred.  Last, in the light of insights drawn 
from the simulations, section 6 examines interpretive options relating the new 
observations of carbon star kinematics to data for other classes of objects, 
principally the neutral hydrogen, and from these, explores alternative 
scenarios for non-gravitational dynamical processes affecting the tidal 
interpretation in the line of sight.  The study ends with an exploration of 
some astronomical implications.

\section{OBSERVATIONS}

The observations were obtained between 1991 and 1997 with the modular 
spectrograph at the duPont 2.5 meter telescope during 42 observing nights, the 
preponderant majority of the data being taken at a resolution FWHM of 1.1\AA\  
over a spectral region between 7850 and 8760\AA, reported earlier (Kunkel, 
Demers \& Irwin 1997, hereafter Paper II). Apart from two initial survey 
fields selected especially for the original discovery mission, the complete 
SMC survey searched 12 ESO/SERC fields: 13, 14, 28, 29, 30, 31, 50, 51, 52, 
53, 79, and 80, with kinematic data for a total of 150 carbon stars redder 
than ($\it B_j - R_f$) = 2.4. Some material not published in the earlier list 
is here included in an appendix.  The data processing consisted in the 
extraction of one-dimensional spectra employing the traditional IRAF data 
reduction packages. After numerous trials with a variety of other template 
spectra, cross-correlation algorithm matching a radial velocity template 
derived from the R-type carbon star HD16115 with a heliocentric radial 
velocity of 14.0 km s$^{-1}$ (McClure {\it et al.~}1985) was found to give the 
most satisfactory reduction of the entire data set.  The master template was 
masked to employ three separate spectral regions known to be almost entirely 
free of telluric features. Two regions covered the CN bands at 7915 and 
8165\AA, avoiding the strong telluric water features near 8165\AA, and a third 
region covering the Ca II triplet between 8550 and 8710\AA. Comparison 
exposures of an argon lamp were taken after each spectrum, to assure that 
velocity errors arising from instrumental flexure could be held to less than 2 
km s$^{-1}$.

Most of the spectra, logarithmically re-sampled, attained a raw signal to noise
ratio of eight or more.  These produced radial velocity estimates that repeated
reliably, and for those stars believed to be free of binary orbital motion, to
within 5 km s$^{-1}$ of their mean (see Kunkel, Demers \& Irwin (1997) for 
details). Increasing the instrumental dispersion yielded no improvement in the 
velocity precision, and we believe that the limiting precision reflects the 
atmospheric motions in these late type atmospheres, as had already been noted 
in the spectrographic data obtained for long period variables (Hughes, Wood, 
\& Reid 1991).

To assure a proper match of the instrumental velocity system between the 
various nights, as well as between data obtained by other observers, 6 or more 
carbon stars were re-observed on most nights, and additional carbon stars were 
included to permit comparison with the extensive observations of Hardy, 
Suntzeff \& Azzopardi (1989). Likewise 16 carbon stars were observed in 5 LMC 
globular clusters for which radial velocity measures had been obtained by 
Olszewski \etal (1991), and 7 stars in 4 SMC clusters for which velocities 
have been reported by Da Costa \& Hatzidimitriou (1998). Since SMC and LMC 
candidates were observed on the same nights, the velocity systems for our data 
are the same for both Clouds.  Stars the velocities of which appeared to 
deviate exceptionally from the population mean were re-observed during several 
observing seasons to confirm their exceptional values, or to justify their 
rejection as radial velocity variables.

\section{SUPPLEMENTARY DATA}

\subsection{\it Carbon stars}
Although our data cover the outer periphery of the SMC completely, we have
supplemented the inner portion of the SMC with data taken from the lists of
Hatzidimitriou \etal(1997) that provides velocities for 72 stars, of which 
there are 19 in common with our survey.  The small proportion of common 
detections is due to different detection criteria employed by those authors, 
resulting in the inclusion of a number of hotter carbon stars than the 
threshold adopted for our work.  An offset of $+9$ km s$^{-1}$ must be added to 
our data to bring them into line with Hatzidimitriou \etal(1998). We included 
6 stars from the list of Hardy, Suntzeff \& Azzopardi (1989) in the entire 
observing program for the purpose of matching our and their kinematic data. 
Jointly this combination forms the data set analyzed in the remainder of this
paper. An offset of $+7$ km s$^{-1}$ added to our data bring them into line with 
those of Hardy, Suntzeff \& Azzopardi (1989). An offset of $+5$ km s$^{-1}$ added 
to our data brings our data into line with the cluster data of Olszewski et al 
(1991); 16 stars in these clusters (NGC1751, NGC1783, NGC1806, NGC1846, and 
NGC2213) were observed during the program covering the SMC carbon stars.

\subsection{\it Globular clusters}
Radial velocity data for seven globular clusters are taken from the 
investigation of Da Costa \& Hatzidimitriou (1998) to which an offset of 
$+18$ km s$^{-1}$ has been added to bring them into line with the two objects 
in common with our data.

\subsection{\it Early type stars}
Spectroscopy of early type stars is from Grondin, Demers, \& Irwin (1992),
covering ESO/SERC fields 31, 32, and 53.  Radial velocities are taken from the 
same source.  Distances to the associations of early type were taken from 
{\it B, V} photometry of Demers \& Battinelli (1998).

\subsection{\it Neutral hydrogen}
Although a number of newer surveys covering the neutral hydrogen distribution
in the ICR are becoming available (for example Putnam \etal 1998), we have 
relied on the survey of McGee \& Newton (1986) primarily to take advantage of 
the smoothing inherent in the lower beam resolution used.  The numerical data 
we abstract employs all Gaussian peaks in their Table 4 greater than 
3.5$\times 10^{19}$cm$^{-2}$

\section{\bf PROJECTION OF THE DATA}
An interpretation of the data is facilitated through specialized projections of
the observational data.  A schematic representation of the orientation of the
orbital plane of the SMC with respect to the SMC/LMC barycenter is depicted in
Figure 1, lying nearly on the observer's line of sight.

\begin{figure}
\centerline{\hbox{\psfig{figure=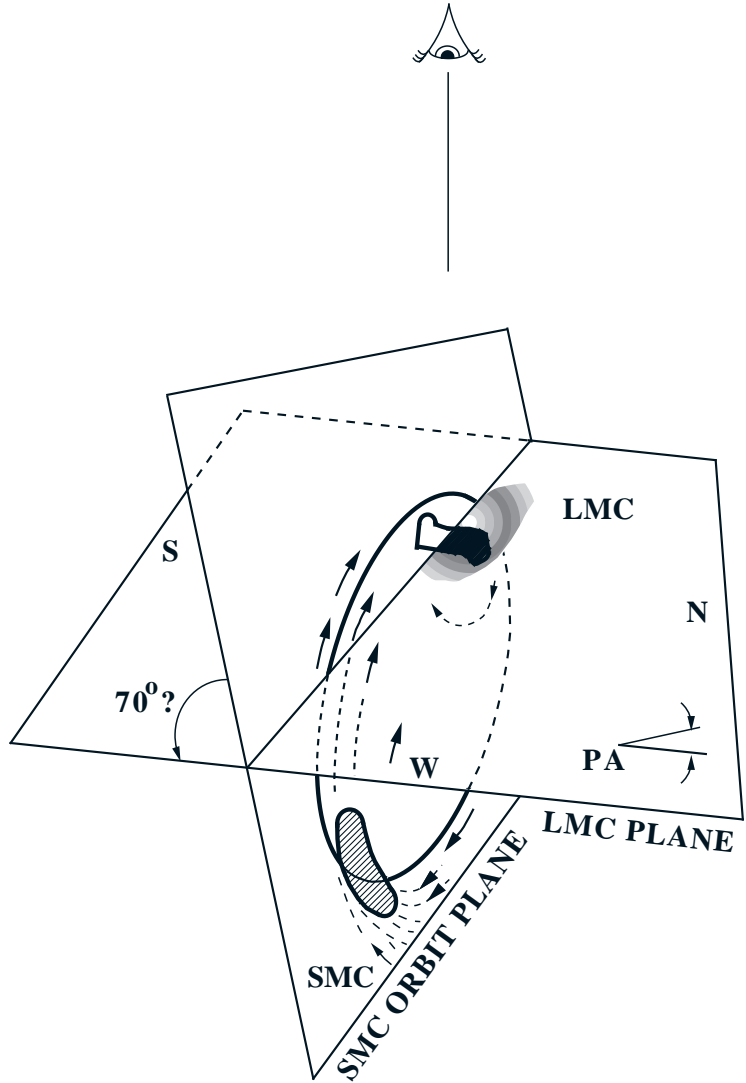,width=0.6\textwidth}}}
\figcaption{
Schematic of the LMC disk plane and the SMC orbital plane about the
          LMC.  Celestial North is to the right, West is toward the reader.
The SMC orbital plane is inclined by 17$^\circ$ to the line of sight, with the
edge nearer the observer lying southward.  In this configuration the LMC
rotates clockwise, and the SMC moved upward in its orbit, through the LMC plane
toward the observer.  Short arrows show velocities of representative debris
particles. 
\label{fig1}}
\end{figure}

A map of the observational data points projected on the plane of the sky is 
shown in Figure 2, with triangles, squares, and stars representing carbon
stars, globular clusters, and stars of early spectral type, respectively.  Two
``clusters'' of tiny open circles near the center of the SMC represent the 
carbon stars observed by Hardy, Suntzeff \& Azzopardi (1989) that form a 
reference population with which our periphery objects are compared. Their role 
in this investigation will become apparent in section 6. Completeness in the 
carbon star data coverage extends to a radius of 12$^\circ$ for the carbon 
stars, and is less for the stars of early type, surveyed in but three ICR 
Schmidt fields between the Clouds. It is also less for all of the 
supplementary data utilized.  Before the observed radial velocities can be 
employed in dynamical interpretation it is important to note that a component 
of proper motion of the SMC in the plane of the sky entails a component of 
velocity along the line of sight that varies with the apparent longitude in 
the SMC's orbit, following the manner first described by Feitzinger et al 
(1977).  Radial velocities employed in the analysis described below have been 
adjusted in this sense, assuming a proper motion parallel to that of the LMC, 
but less by some 60 km s$^{-1}$ (Irwin \etal 1996).  From the proper motion 
measured in our study of the LMC carbon stars (Kunkel \etal 1997, hereafter
Paper III) we adopt a value of 180 km s$^{-1}$ at PA($\it l,b$) = 310$^\circ$.


\begin{figure}
\centerline{\hbox{\psfig{figure=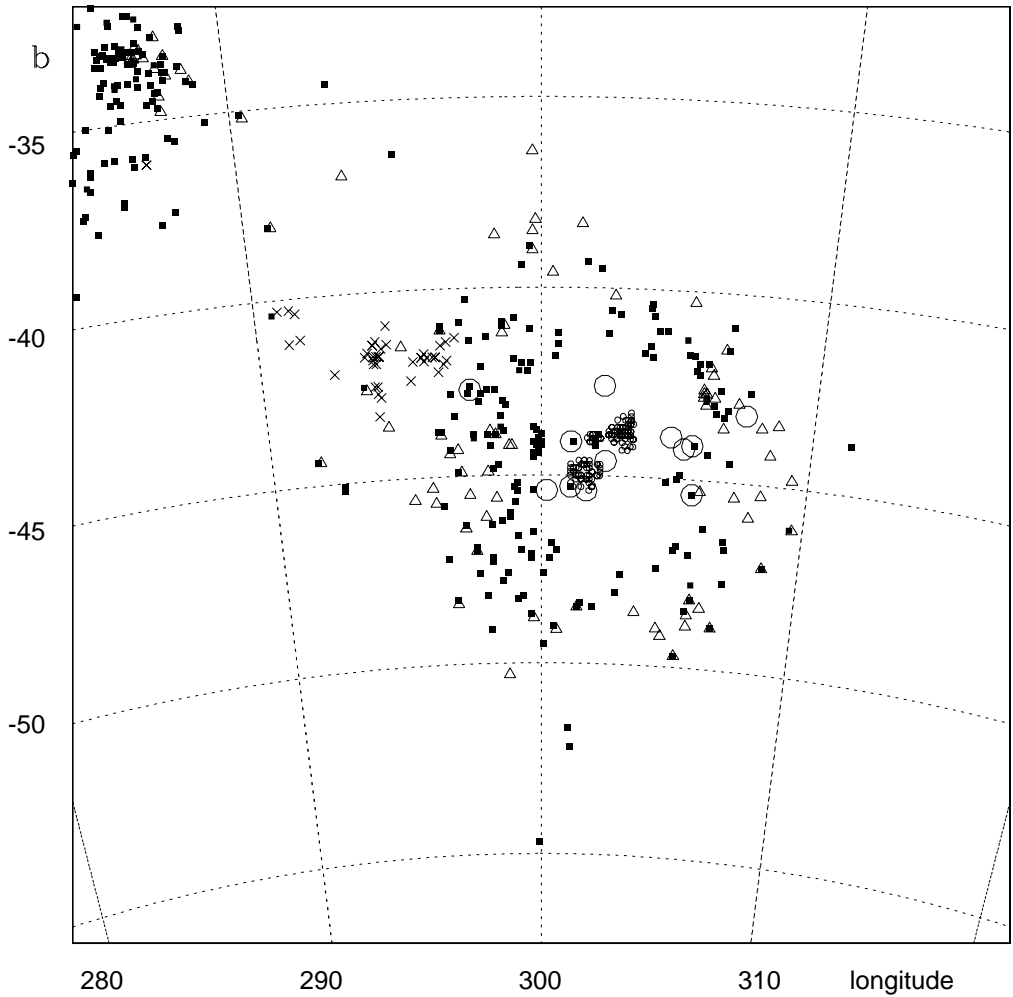,width=0.9\textwidth}}}
\figcaption{
A projection on the sky of the carbon stars 
observed for this investigation (filled squares), the carbon stars observed 
by Hatzidimitriou \etal (1997) (open triangles), the carbon stars of Hardy, 
Suntzeff \& Azzopardi (1989) (small open dots ``caviar''). Star symbols are 
stars of early spectral type. Globular clusters are shown as large open
circles.
\label{fig2}}
\end{figure}


\begin{figure}
\centerline{\hbox{\psfig{figure=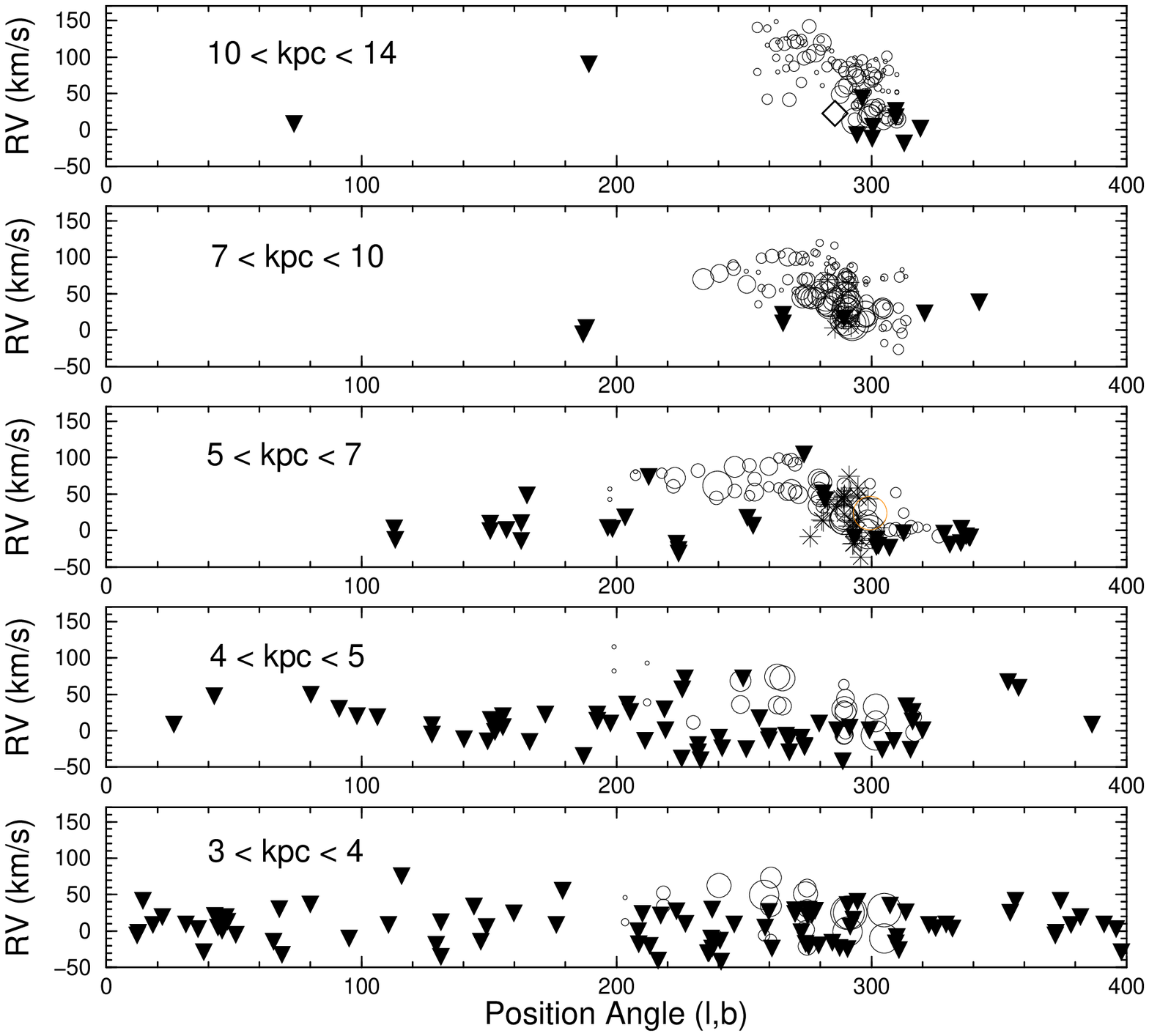,width=0.9\textwidth}}}
\figcaption{
Radial velocity of carbon stars (filled triangles), early type stars (stars), 
and samples of HI (open circles) from McGee \& Newton (1986) are shown as a 
function of position angle about the SMC photocenter (ref) in 5 annular zones, 
from bottom upward, between 3, 4, 5, 7, 10, and 14 degrees, and as many kpc at 
the adopted SMC distance.
\label{fig3}}
\end{figure}

The anomalous distribution of objects on the periphery of the SMC is readily
apparent in figure 3, which shows a ``panoramic'' display of the distribution
within annular zones in five ranges of radii, centered on the photocenter of 
the SMC (at $\it l,b$ = 302.8$^\circ$, --44.3$^\circ$), as a function of 
position angle (zero toward the North Galactic pole).  From the bottom toward 
the top the 5 zones lie between radii of 3, 4, 5, 7, 10, and 14 degrees, and 
as many kpc at the adopted distance of the SMC.  The areas covered by the 
annular zones are approximately 16, 28, 75, 150, and 300 square degrees of 
sky. The direction toward the center of the LMC is toward PA = 294.5$^\circ$; 
the celestial North pole lies toward PA = 182$^\circ$. East is toward PA = 
272$^\circ$. The lowest of the 5 annular zones shows a fairly uniform 
distribution of carbon stars, with no measurable component of rotation. Open 
circles between 200$^\circ$ and 320$^\circ$ of PA represent H I samples from 
McGee \& Newton (1986), the smallest symbols corresponding to 
3$\times 10^{19}$ cm$^{-2}$, and the largest 80$\times 10^{19}$ cm$^{-2}$.  As 
we range outward beyond 4 kpc the second panel from the bottom shows a 
tendency for carbon stars to thin out in the PA range between 320$^\circ$ and 
80$^\circ$, to the SW of the SMC, and opposite of the direction toward the 
LMC. At radii beyond 5$^\circ$, carbon stars are found primarily to the East 
of the SMC, with a slight preference southward of the line toward the LMC.  At 
larger radii carbon stars are seen concentrated along a line directed toward 
the LMC (toward PA = 295$^\circ$). In the outermost zone a group of 9 ICR 
carbon stars group within the range 292$^\circ <$ PA $< 320^\circ$; two
additional carbon stars at lesser PA's are most likely unrelated interlopers
one expects from the field density reported by Totten \& Irwin (1998).  The
trend concentrating ICR carbon stars toward the great circle connecting the
Magellanic Clouds, as well as the decrease in density (count per square degree)
at larger separation from the SMC, we interpret as demonstrating the SMC to be
the origin of the ICR carbon stars, rather than the LMC.
\vskip 0.3cm
The upper three panels of Figure 3, expanding the view toward the LMC, offer
interesting dynamical information that we examine in greater detail in figures
4 and 5.  In the lower panel of Figure 4, in the range of PA's between 
270$^\circ$ and 300$^\circ$, the H I samples of greatest density are seen to 
lie along a spine whose velocity decreases from about +70 km s$^{-1}$ at PA = 
270$^\circ$ to 0 km s$^{-1}$ at PA = 300$^\circ$. The upper panel of the 
figure shows much the same negative slope, though a bit steeper, from +100 
km s$^{-1}$ at PA = 270$^\circ$ to +15 km s$^{-1}$ at PA = 300$^\circ$.  The 
upper envelope of the HI likewise conforms to a similar slope, and, though 
there are too few carbon stars to convince by themselves, what positive 
velocity carbon stars appear in the lower panel follow the same trend.


\begin{figure}
\centerline{\hbox{\psfig{figure=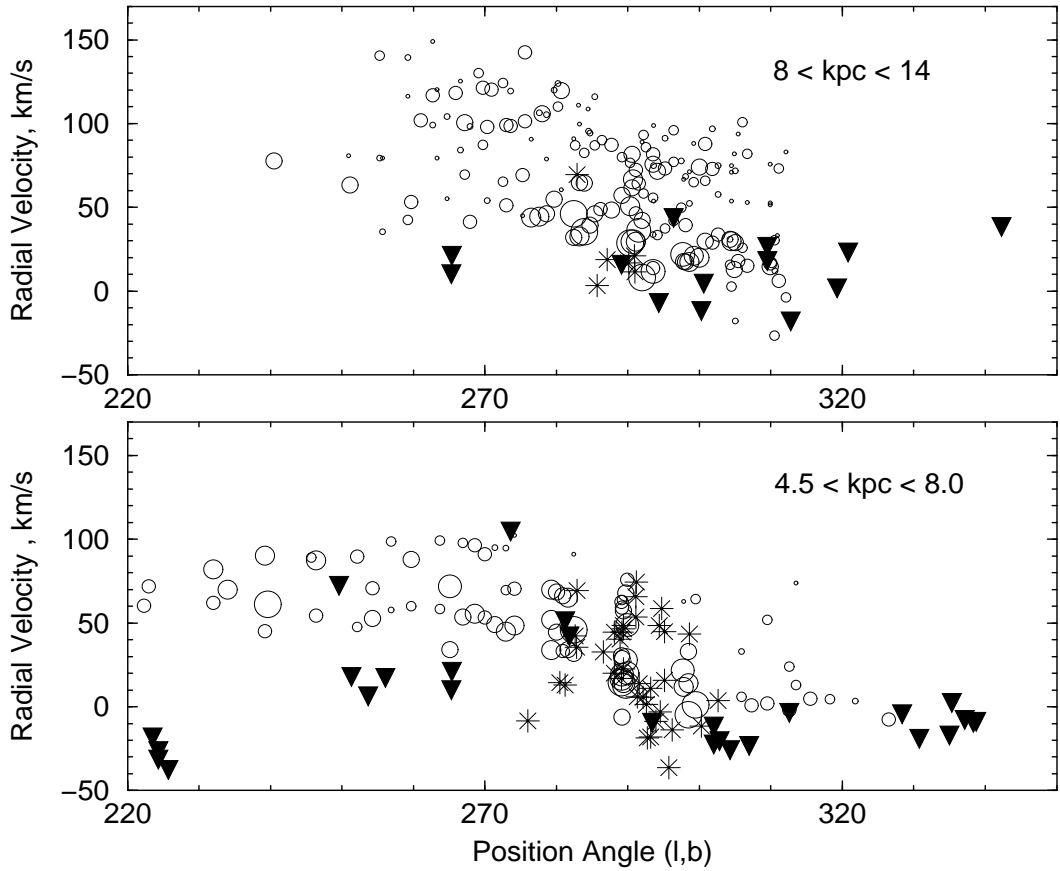,width=0.85\textwidth}}}
\figcaption{
Panoramic view of radial velocities in two annular zones, toward the direction 
of the LMC (PA = 295$^\circ$) of carbon stars (filled triangles), early type 
stars (stars), and samples of H I (open circles) as seen from the SMC 
photocenter.
\label{fig4}}
\end{figure}

An estimate for the inclination of the SMC orbital plane to the line of sight  
may be constructed from this gradient in the H I spine and the gradient of the
observed radial velocities of the red horizontal branch clump stars in the line
of sight, using the data from the eastern edge of the SMC bar (Hatzidimitriou,
Cannon \& Hawkins 1993).  Their velocity gradient of 70 km s$^{-1}$ over a 
10 kpc range of distance in the line of sight provides scaling via a 
transformation between velocity and distance. The energy of motion in the line 
of sight certainly precludes the continuation of the SMC as a single 
self-gravitating entity. Hence we adopt an approximate scaling between radial 
velocity and distance.  In the absence of significant rotation in the SMC, and 
assuming that ICR particles conserve their angular momentum about the LMC/SMC 
barycenter, two H I samples in the lower panel of Figure 4 separated by 70 
km s$^{-1}$ place the more positive sample some 10 kpc more distant.  This 
difference in velocity is seen between the densest samples at PA = 270$^\circ$ 
and similarly the densest samples at PA = 300$^\circ$, whose separation in the 
plane of the sky is about 3 kpc.  Consequently the line of sight lies at an 
inclination of about arctan(3/10) = 17$^\circ$ at most. Comparison of the data 
with simulations described in section 5 suggests that the angle is somewhat 
less than this.

When comparing the carbon stars in the ICR to the neutral hydrogen data of
McGee \& Newton (1986), the distribution of the carbons in the ICR is seen at 
generally more negative velocities for any position angle or distance from the 
SMC, almost forming the low velocity envelope, as is apparent in both panels 
of Figure 4. Moreover, the most positive velocities are found to just one side 
of the direction toward the LMC: the northern (PA $<$ 295$^\circ$). The 
positive velocity domain is not devoid of carbon stars.  Two are shown to lie 
at the upper envelope of the H I.  Figure 5 demonstrates the effect in a pair 
of ``fan'' plots sampling two triangular areas spaced symmetrically about the 
direction toward the LMC.  The fans show radial velocities of particles lying 
within a narrow range of position angles seen from the SMC photocenter, as a 
function of radial distance from the SMC photocenter toward the LMC. The North 
fan (250$^\circ <$ PA $< 290^\circ$) is dominated by neutral gas (open circles 
with diameters corresponding monotonically to H I surface density), containing 
but three carbon stars in the range between 7 and 14 kpc, while in the same 
distance range the South fan (300$^\circ <$ PA $< 340^\circ$) contains 7 
carbon stars. Interestingly, the radial velocities of the carbon stars in both 
fans lie within or at the lower velocity envelope defined by the gas.  Other 
traits distinguishing the two fans are (1) the comparative dearth of neutral 
hydrogen in the South fan, and (2) the pronounced dense clump of carbon stars 
in the South fan in the radial range 14.8 $<$ kpc $<$ 16.9 only weakly seen in 
the North fan.


\begin{figure}
\centerline{\hbox{\psfig{figure=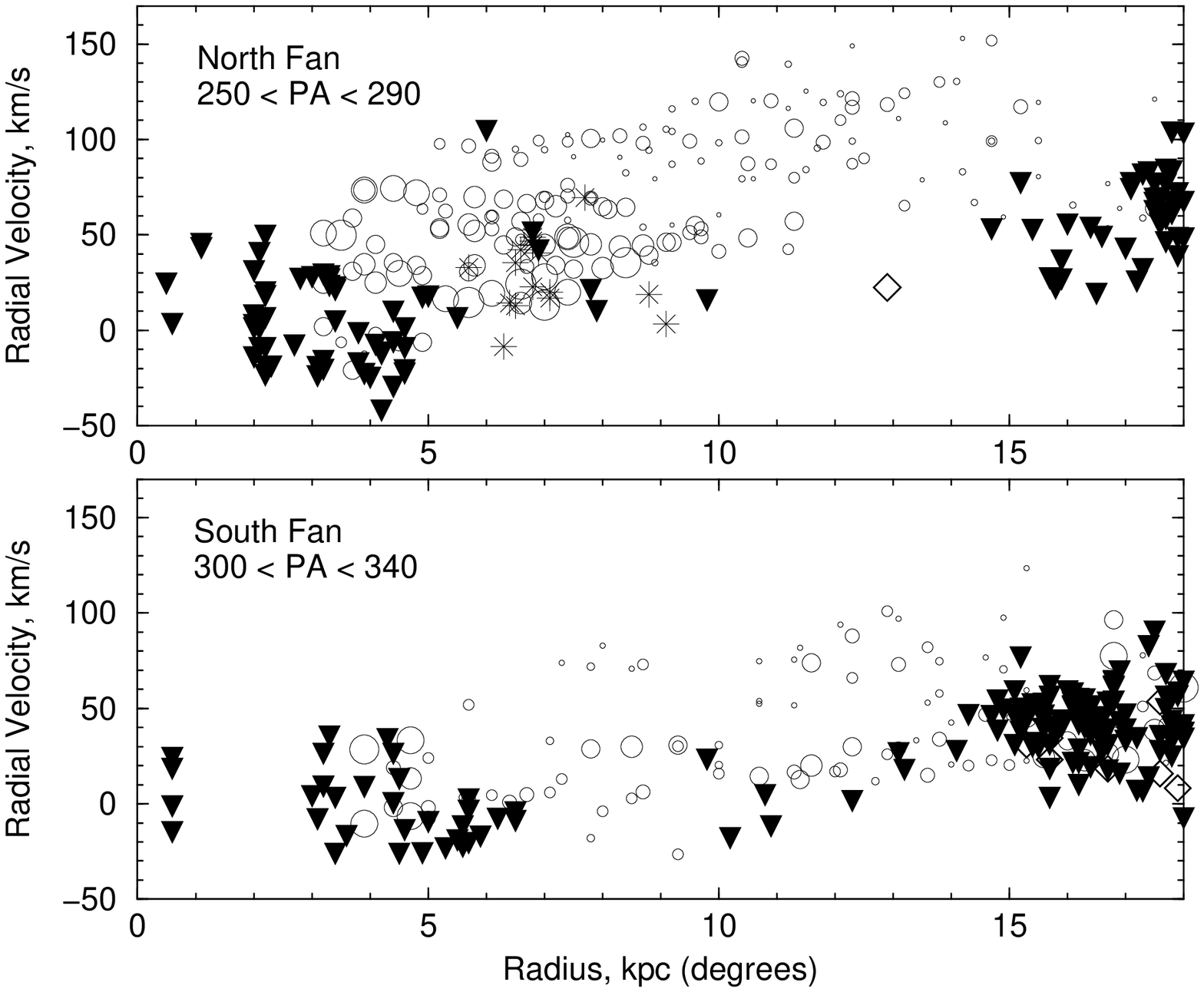,width=0.85\textwidth}}}
\figcaption{
``Fan'' plots showing radial velocities of particles (symbols as in
          figure 4) as a function of radial distance from the SMC photocenter.
The upper panel includes particles lying at PA's between 5 and 45 northward of
the direction to the LMC, and the lower panel shows particles in the fan between
5 and 45 degrees south of the direction toward the LMC.
\label{fig5}}
\end{figure}

Several interpretations are implied by these observations.  First, with the
position of the SMC behind the LMC (and hence behind the LMC/SMC barycenter)
orbital motion places the South fan closer to the observer, and similarly,
places the bulk of the gas at distances greater than the ICR carbon stars.
Furthermore, though less obvious, we note that in those portions of the
diagrams where there is a separation between carbon stars and gas, that
separation is consistently less when comparison is with the more massive
(denser) H I samples. It is as though the mechanism segregating the carbon 
stars from the gas were tuned to density, affecting the denser H I clouds in 
the same way.  It is worth re-stating this finding in a converse sense: that 
the particles most affected by the separating mechanism are the H I samples of
lowest density.  Last, we attach importance to the fact that whatever
mechanism has depleted gas from the forward edge of the orbiting SMC gas, even
the most remotely trailing gas is not totally devoid of stars, as the two
carbon stars in the lower panel of figure 4, at (RV, PA) = (71 km s$^{-1}$, 
249$^\circ$) and (106 km s$^{-1}$, 273$^\circ$) demonstrate. The 
interpretation of these features of the data is delayed to section 6.

Stars of early spectral type are seen superposed preferentially on the denser 
agglomorations of H I, with a slight preference for the more negative 
velocities. In all, the inhomogeneous distribution of the debris material 
suggests that the side nearest the observer, the low velocity edge, has 
experienced a significant depletion of the neutral hydrogen, since otherwise, 
with gravity alone acting, the ratio of carbon stars to gas should not vary 
with velocity as is seen to prevail over most of the ICR.

In closing this section we note that our interpretation of the sense of
orbital motion poses no conflict with the orientation of the SMC bar reported
by Mathewson, Ford, \& Visvanathan (1986, 1988) and Hatzidimitriou \& Hawkins 
(1989) and Hatzidimitriou, Cannon, \& Hawkins (1993) which places the NE end of
the bar closer to us.  This apparent discrepancy merely indicates that the SMC
bar does not move parallel to its orientation in space, but ``yaws'' southward,
much as the orientation of an aircraft is seen to ``yaw'' with respect to its
flight direction in a crosswind.

\section{INTERPRETATION}
The search for an interpretive paradigm relies on dynamical simulation using a
simple n-particle ``PP code'' in which the LMC and SMC are represented by
approximately 1500 particles each, allowing them to adjust their potentials
under self-gravitation.  The purpose of such a simulation is to guide the
search for orbital and structural parameters describing each of the Clouds and
the details of their interaction, in which discrepancies between a simulation
and observations can serve to (1) seek improvements in the simulation, (2)
guide the observational program during its development, and (3) steer the 
interpretation effort around unsuspected pitfalls, of which the most 
problematic is that of learning what happened to the primordial internal 
angular momentum (spin) of the pre-encounter SMC disk.  The second aspect 
controlled the last two years of the observing program and the first two have 
by now been met.

\subsection{\it The paradigm}
A search for a compact description of the interaction between the Magellanic
Clouds will not be advanced significantly if the simulation is required to
model every and all physical processes one observes or might consider
interesting, including for example the star formation precipitated by a tidal
impulse, and which we see unevenly distributed over the face of the SMC.  The
approach preferred here seeks to {\em minimize} the number of parameters 
needed to describe the essentials of the gravitational interaction alone 
between particles that experience no internal evolution during the time 
interval studied, and to exclude other forces, such as those arising through
hydrodynamics or from magnetic fields.  Some detail important to local
astronomical processes will be lost, such as the distribution of star formation
regions or the energy input to the ISM from supernova events.  We call this
approach a minimum parameter technique (MPT).  Its principal aims are (1) to
test the adequacy of interpreting a tidal disruption as dominated by gravity
alone on visible particles only, and (2) to indicate something of the character
of additional physical mechanisms (or particles) that may affect the kinematics
observed.

The motivation for preferring the minimization of the parameter count is not 
merely philosophical but is rooted in the nature of the mathematical character 
of our search.  In practice, even at the outset, a large proportion of the 
parameters controlling a simulation can be narrowly constrained, such as the 
coordinates of the major perturbers in phase space.  If all but a very few 
parameters can be specified in advance, the simulation should rapidly achieve 
its goals. However, as the count $k_f$ of free parameters increases, the 
volume of neighboring parameter space that must be searched at each 
integration step rises explosively, the effort to search the neighboring 
pieces of parameter space increasing roughly as 2$^{k_f}$ at each occasion.  
Such a search cannot be expected to converge speedily to the description 
sought, however. In a worst case 2$^{k_f}$ simulations are required before a 
succeeding iterative step in parameter space may be chosen optimally.  And 
worse still, there is nothing but faith to sustain the belief that the 
topology of the immediate neighborhood in this parameter space is well behaved 
(continuous and differentiable), so that iterative methods stand a chance of 
converging.  Our experience has shown that at least one parameter is formally 
chaotic: disk structures are affected by the Toomre instability to bar 
formation.  Moreover, our methodological simplification introduces other 
difficulties in choosing the next time step. In advancing a simulation step 
truncation errors sum, and the graininess inherent in the small count of 
particles requires control, and we have followed the guidelines recommended by
Merritt (1996).  For example, a minimum particle count was determined running 
simulations with different particle counts and observing the precision with 
which descriptive detail settled to a unique configuration at some minimum 
value;  runs with at least double, and usually five times such minimum values 
were used to constrain the buildup of error.  In effect, beyond the summing of 
errors, the topology of parameter space has become fractal.  In practical 
terms, the significance is that there is no assurance that iterative methods 
can explore the entire parameter space one would like to dominate.  Simple 
iterative strategies can explore only a small fraction of parameter space, and 
this limitation forces us to adopt a compromise:  By minimizing the count of
parameters some cosmetic detail in the simulation may appear deficient in
describing the observations.  However, the investigator may decide later, after
a number of iterative improvements have been achieved, if the acceptance of
such concessions has derailed the exercise.  It is the fractal character of
parameter space that leads to the following caveat:  just because a simulation
has been found that matches an aspect of the observations for some interesting
combination of parameter values one has no assurance that the neighborhood of
the solution judged as acceptable is multiple or single valued, and whether
other, physically more meaningful parameter combinations may lie nearby.
Historical examples of such circumstances now merit review, and a few will be
briefly mentioned further on.  An anonymous precaution summarizes this dilemma
and is worth keeping in mind: ``all simulations are doomed to succeed!'' (Welch,
1998), and increasing the parameter count beyond an MPT aggravates this
situation.

\subsection{\it Parameter choices}

The simulation shall describe the interaction of the Magellanic Clouds in their
motion about the Milky Way.  More remote perturbers, such as M31, are not
treated: as without significance during the 1.2 Gyr duration of the simulations.
There are 26 parameters that are deemed to constitute a minimal configuration.
Their initial values (count in parentheses), and their search ranges are:

\begin{enumerate}
\vskip 0.3cm
\item The mass ratios (2; 1 free)
\vskip 0.3cm
\begin{enumerate}
\item     ${\cal M}_{SMC}/{\cal M}_{LMC} = 0.25$
\vskip 0.3cm
\item 0.007 $<$ ${\cal M}_{SMC + LMC}/{\cal M}_{Galaxy}$ $<$ 0.05
\end{enumerate}
\vskip 0.3cm
\item Phase space (12; 1 free)
\vskip 0.3cm
\begin{enumerate}
\item     Positional coordinates are adopted, setting radial distances as 50 
          and 58 kpc for the LMC and SMC, respectively.
\vskip 0.3cm
\item     For both Clouds the velocity in the line of sight is adopted from 
          the observations. For both Clouds the direction of motion in the 
          plane of the sky is taken to be that of the Magellanic Stream.
\vskip 0.3cm
\item     The speed of the LMC can be solved for by the method of Feitzinger 
          \etal (1977): we adopt that transverse speed for which the line of 
          nodes retains the least warping (= 240 km s$^{-1}$), which is close
          to the 215 km s$^{-1}$ obtained by Jones, Klemola and Lin (1994) for
          an LMC mass closest to our earlier value (Paper III). For the SMC we 
          adopt a  transverse speed 60 km s$^{-1}$ less, in  conformity to the 
          geometric and timing constraints (since pericentric passage).
\vskip 0.3cm
\item     The motion of the SMC perpendicular to the Magellanic Stream is
          constrained by the inclination of the SMC orbit to the plane of the 
          sky (section 4 above).
\end{enumerate}
\vskip 0.3cm
\item  Spin orientation (6; 2 free)
\vskip 0.3cm
\begin{enumerate}
\item     The spin of the of the Milky Way is relatively unimportant for the 
          simulation.  Motion of the LSR was set to 225 km s$^{-1}$.
\vskip 0.3cm
\item     An inclination of 33$^\circ$ and line of nodes for the LMC is adopted 
          from Feitzinger et al. (1977).
\vskip 0.3cm
\item     There is no indication in the observations of the SMC that any
          meaningful tracer of spin information is preserved;  in comparison 
          to other motions the enormous extent of the SMC in the line of 
          sight suggests that there may no longer be any significant spin. 
          In many ways this is the most demanding constraint on the 
          simulation, and especially for a description of the SMC core.
\end{enumerate}
\vskip 0.3cm

\item  Structure (6; 3 free)
\begin{enumerate}
\item     The question of the number of parameters describing the disk and/or
          halo structure is the most debated among aficionados of simulations.
          In conformity with the MPT approach a minimum choice must include a
          scale length, a scale height, and the fraction of the total energy of
          internal motion that shall be contained in particle heat (instead of
          cold rotation).  The last two quantities are not independent.
          Stability of the disk against bar formation is controlled by Toomre's
          Q.  No halo component as a separate entity is provided; its separate
          inclusion would augment the parameter count by six, requiring too
          much arbitrariness in devising an iteration strategy.  Furthermore,
          since tidal action must affect dark and luminous matter similarly, we
          cannot find [devise] a rationale for treating their dynamics
          separately.
\vskip 0.3cm
\item     We adopt an LMC scale length of 1.7 kpc (Feitzinger 1980).  We adopt
          (arbitrarily) an axis ratio (scale height) of 0.3, and a velocity 
          dispersion (heating) to bring the quiescent disk to the point of bar 
          instability.  The resulting particle dispersion in the simulation
          amounted to 19 km s$^{-1}$, not far from the 16 km s$^{-1}$ observed
          in the LMC carbon stars (Kunkel \etal 1997).
\vskip 0.3cm
\item     There is no indication about what structure the SMC may have had
          prior to its encounter with the LMC.  A relatively cool disk similar 
          to that adopted for the LMC appears unlikely;  in the absence of a 
          pre-encounter SMC bar all simulations show that following an 
          interaction a significant portion of internal spin should have been 
          preserved, and no evidence for such spin is found in any observations 
          of the SMC bar. Far greater disruptive damage to the SMC periphery 
          would be required than that observed to mask the residual spin.  In 
          view of the fractal character of parameter space, we consider that 
          several choices of a pre-encounter configuration produce equally 
          valid SMC Wing structures. The structures so far explored are either 
          an essentially spin-free spheroidal configuration reminiscent of the 
          dSph systems, or, what seems more likely in view of the high
          incidence of internal bars Odewahn (1994) finds in Magellanic
          satellites, a strongly bar-like system with the bar orientation at 
          pericentric passage along a line of strongest tidal shear.
\end{enumerate}
\end{enumerate}

\subsection{\it Summary of the simulations}
Of the 26 parameters selected, 7 are considered ``free'', in that they are
allowed to vary over a reasonable range, permitting some adjustment of the
match between the simulation and the observational data.  Some parameters
appear quite ``robust'', varying independently of the others.  Most noteworthy
in this sense is the total Magellanic Clouds-to-Galaxy mass ratio,
${\cal M}_{SMC+LMC}/{\cal M}_{Galaxy}$. Within the tested range the value of 
this quantity was found narrowly constrained to lie between 0.020 and 0.030, 
and will be more carefully discussed in a later paper dealing with the carbon 
stars of the Large Magellanic Cloud.  


The free phase space parameter (of 12) is the proper motion of the SMC
southward with respect to the SMC,LMC pericenter. The rationale confining this
quantity was explored in section 4.0 above, and is basically determined by the
discussion surrounding Figures 3 and 4.

The two parameters governing the spin orientation and the three disk structural
parameters were the outstanding headache dominating the selection of simulation
parameters.  The principal difficulty consists in the fact that any significant
spin in the pre-encounter SMC disk, if present, survives the encounter to
excessive degree for all reasonable orbital impact parameters at the current
SMC/LMC encounter.  Adding a ``massive dark matter halo'' on a scale other than
``mass follows light'' to the SMC or LMC acts in the wrong sense, aggravating 
the dilemma.  In fact, the most satisfactory ``fits'' to the observed 
distribution of SMC material were obtained by leaving the pre-encounter SMC 
virtually devoid of any pre-encounter spin at all!  The most influential 
parameter affecting the ``goodness of fit'' was found to be the Toomre Q, and 
for the small particle count in our simulation that parameter behaves formally 
in a chaotic manner.  In practical terms, we found that when the Toomre Q was 
set close to its critical value, even an infinitesimal variation in some other 
parameter affecting the SMC/LMC encounter (such as a small change in the 
particle count!) elicited major variations in the resulting debris 
distribution.  Our primary conclusion regarding the remaining five ``free'' 
parameters is that, were one to assume a typical disk structure as observed in 
other dwarf disk systems of comparable luminosity, then our simulations were 
unable to find a configuration with a spin amplitude typical of what one 
associates with a dwarf galaxy of the SMC's luminosity.  Alternatively, were 
the pre-encounter SMC endowed with a significant bar, then the orientation of 
that bar during the encounter often dominates the later debris distribution.  
Even then, however, any ``reasonable'' spin attributed to the SMC spreads 
debris over too wide a swath between the Clouds, unless one reduces the 
SMC/LMC mass-ratio, as was done by Fujimoto \& Sofue (1976, 1977), Murai \& 
Fujimoto (1980) and in the later simulations of Gardiner \& Noguchi (1996).  
In summary: a good fit to the SMC Bar and Wing can be attained for a varied 
choice of structural parameters, most easily by keeping the SMC's spin low and 
``leaning'' the pre-encounter bar to the ``right'' angle.  The vagaries of the 
Toomre Q limit the effectiveness of such a search, and finding one satisfying 
configuration does not signify that others, with equally valid parameter 
values, may not exist. We believe the solution to this problem to be 
multi-valued, so that without additional observational data any one successful 
simulation tells little of what actually happened.

The bottom line of this inquiry may well be that one must treat the role of
massive dark matter haloes with care if the absence of SMC core spin is to be
adequately represented.

\section{\bf CORROBORATING DATA IN THE LINE OF SIGHT}
A demonstration that tidal disruption has affected the SMC throughout its
entire volume may be appreciated from component population inhomogeneities in
the line of sight (LOS) toward the photocenter of the SMC where no visible
effects appear in direct imaging.  This demonstration depends on the reasoning
employed to determine the SMC's orbital inclination in section 4 above.  If we
accept that the SMC is currently experiencing a breakup in the LOS direction,
as has been suggested by the distribution of Cepheid variables studied by
Mathewson, Ford \& Visvanathan (1986, 1988), and more clearly in Red
Horizontal Branch stars on the eastern edge of the SMC bar (Hatzidimitriou \&
Hawkins 1989, Hatzidimitriou, Cannon, \& Hawkins 1993), then it becomes
reasonable to ask whether, during such a breakup of the SMC, the radial
velocity of any SMC object also reflects, at least in part, the position of
such an object in the line of sight.  More specifically, for such objects known
to show low velocity dispersions in the LMC, Hatzidimitriou (1998) finds
significantly the higher velocity dispersions in the SMC among all classes of
SMC objects investigated, pointing to the presence of an expansion component
that all share uniformly.

This circumstance provides an unusual opportunity to conduct a simple
observational test that is entirely independent of any physical or dynamical
model, and in particular, one which avoids all dependence on reddening that
normally plagues investigations of distributions in radial distance.  In this
test, radial velocity is proposed as an indicator of relative distance, with a
scaling adopted from Hatzidimitriou et al (1993) of 7km s$^{-1}$ per kpc.  The 
validity of such relative scaling rests on the conservation of angular 
momentum of the SMC fragments about the SMC/LMC barycenter.  Such a test will 
prove more sensitive for objects known to show the lower velocity dispersion, 
such as H I, stars of early spectral type, cepheids, supergiants, and, yes, 
carbon stars, given their low intrinsic velocity dispersion observed in the LMC 
(Kunkel \etal 1997). In the application proposed here, the test is formulated 
as a null hypothesis which predicts the following:  Under gravity acting alone, 
if no disruptive expansion were taking place, then in the absence of 
inhomogeneity the median radial velocity observed for all classes of objects 
should be the same. Further, half the objects of any class should lie in front 
of, and half behind this median velocity.  Moreover, if the velocity dispersion 
for two classes of objects that are compared is the same, or small compared to 
an expansion, distribution curves for both classes run parallel.  This last 
concept can be stated more conveniently: the ratio of the portion of one class 
lying in front of any point scaled in that class along the line of sight should 
be the same as for the other class, irrespective of distance.  So, for example, 
and object of the first class, lying behind, say, eighty percent of its own 
class should lie behind eighty percent of the other class as well.

A visual demonstration is most readily arranged in some convenient sequential
counting order, using as independent variable the gas fraction (of H I) with a
velocity less than that of the carbon star (i.e., the gas fraction lying in
front of the star), determined from planimetry under the HI velocity profile.  
Carbon stars embedded uniformly in the gas should lie on a straight line of 
unit slope. This choice of independent variable offers several advantages. One
is that ``self-propagating star formation'' and other non-uniformities affecting
the distribution of young objects in the gas are circumvented.  Another is to
test the relation between the SMC and the Magellanic Stream: If, following 
current lore, the Magellanic Stream is devoid of stars, then the cumulative 
distribution of carbon stars should form a straight line broken on the near low 
side if the Stream lies in front of the SMC, or at the high end of the data set 
if it lies behind the SMC.  In either circumstance, with the presence of the 
Stream a slope greater than 1.0 is expected everywhere except at the Stream, 
where a much flatter, or zero slope should prevail.  Again, if velocity 
dispersions of the gas and stars are very dissimilar, so that stars have 
traveled further in the line of sight than the gas, a flat curve is expected, 
with very steep slopes $>>$ 1.0 at both ends, symmetric about the median 
velocity of the gas.  Last, if no breakup is occurring, and velocity is no 
indicator of distance in the line of sight, then perfect mixing should prevail; 
which is to say that a slope of unity is expected over the entire range of the 
distribution.  On the other hand, if radial velocity does reflect relative 
distance, a slope less than 1.0 over a significant portion of the distribution 
signifies a prevalence of H I over the affected portion of the corresponding 
distance scale, and that any explanatory dynamics cannot be attributed to 
gravitational forces alone.


\begin{figure}
\centerline{\hbox{\psfig{figure=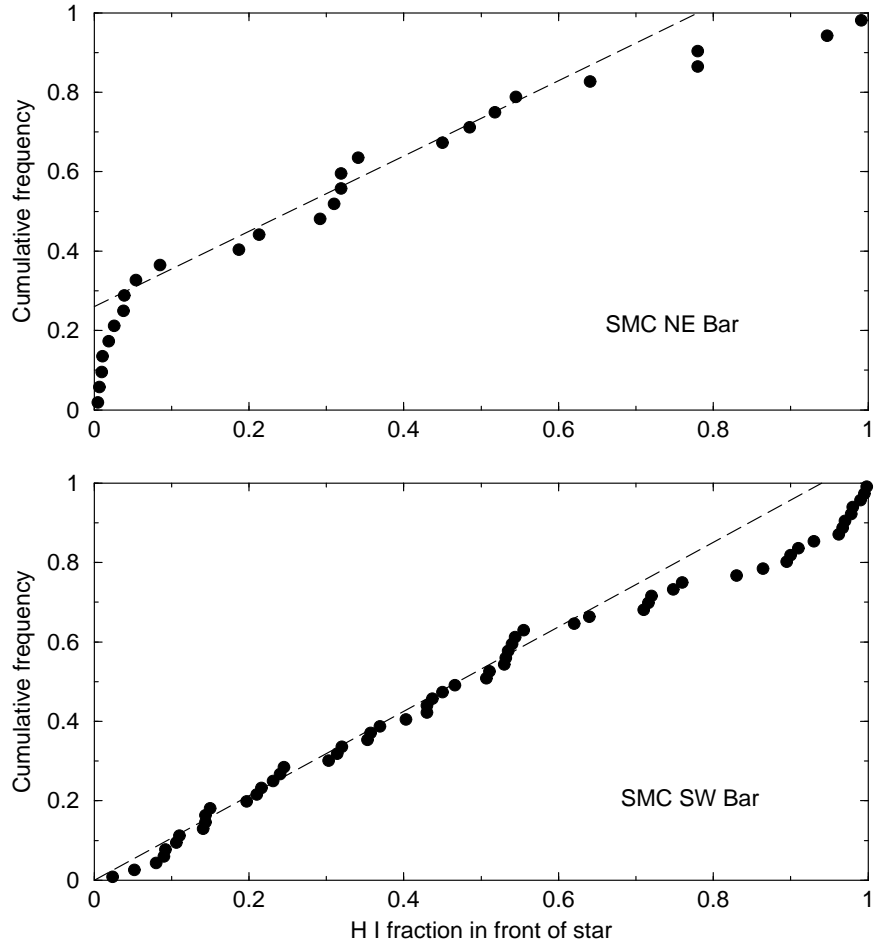,width=0.7\textwidth}}}
\figcaption{
Cumulative distribution of carbon stars (from Hardy, Suntzeff, \& Azzopardi 
1989) as a function of the hydrogen fraction (from Mathewson \etal 1986) lying 
in front of a star. The Southwest Bar is shown in the lower panel, and the 
Northeast Bar is shown in the upper panel.
\label{fig6}}
\end{figure}

A cumulative distribution of the carbon stars from Hardy et al (1989) as a
fraction of the gas (Mathewson \etal 1988) is shown in Figures 6. The
power of this test is predicated on the age of the carbon stars, which is
greater than the time elapsed since the LMC/SMC interaction.  Since the stars
are older, gravitational forces mediating the tidal effects alone act on them,
and deviations from the distribution of the carbon stars necessarily reflect
additional dynamic processes operating on the gas.  The null hypothesis
(absence of additional gas processes) requires a straight line of unit slope,
and it is apparent that for neither the SW bar of the SMC (lower panel of 
Figure 6), nor the NE bar (upper panel) can this condition be said to hold.  
Even in the more uniform sample (the SW panel), drawn from within one degree of 
the photocenter of the SMC, one perceives two slopes, with a break at a point 
two thirds from the front of the SMC.  Compared to the front two thirds of the 
mass distribution mapped by the carbon stars along the line of sight,  for 
which the ratio of gas-to-carbon-stars remains constant, the back third is
quite unevenly mixed, with nearly a factor of two excess of gas between F = 0.64 
and F = 0.84, and almost no H I at all for the last 15 percent of the carbon 
stars.  The situation in the NE bar is more extreme, with the front third of 
carbon stars embedded in less than 5 percent of the H I, and with 70 percent of 
the H I lying behind the midpoint of the carbon star mass distribution. The 
interpretation of these data, taken within one degree of the photocenter of the 
SMC, is that if we are confronted by a break-up of the SMC, then the presence 
of non-gravitational forces modifying the structure of the gaseous component of 
this stellar system must act, or other processes ionizing portions of the gas 
``out of sight.''

The obvious test for the significance of these data is a Kolmogorov-Smirnov
({\it K-S}~) test (Kendall \& Stuart 1961). By this test the probability that 
the observed distribution arises from the same distribution is 11 and 3 
percent for the SW and NE sample, respectively. A description of this test by 
Kendall \& Stuart points out that the {\it K-S} deviation is taken 
irrespective of sign, and irrespective of location of the deviation on the 
abscissa, thereby weakening the discriminative information in the data.  By 
allowing one to select the location of the independent variable, a Monte-Carlo 
simulation of the null hypothesis places far more stringent limits.  In one 
million simulations less than 1.6 percent deviate more than the SW sample in 
the lower panel of Figure 6 at F = 0.88, and less than 0.03 percent deviate 
more than the NE sample in upper panel at F = 0.30.  We note furthermore that 
in the two fields, the deviation from a straight line occurs at opposite ends 
of the line of sight.  In the SW field the deviation occurs at the far end, 
with a greater density of carbon stars compared to HI on the nearer side. That 
is, the gas to stars ratio is greater at the far end of the SMC. Within one 
degree, in the NE field, a dramatically more pronounced excess of stars occurs 
at the front end of the SMC, with 30 percent of the stars lying in front of 95 
percent of the H I! Our conclusion is that the null hypothesis is rejected 
with a confidence exceeding 99.9 percent, and forces other than gravity alone 
are required to explain the kinematics of the SMC.


\begin{figure}
\centerline{\hbox{\psfig{figure=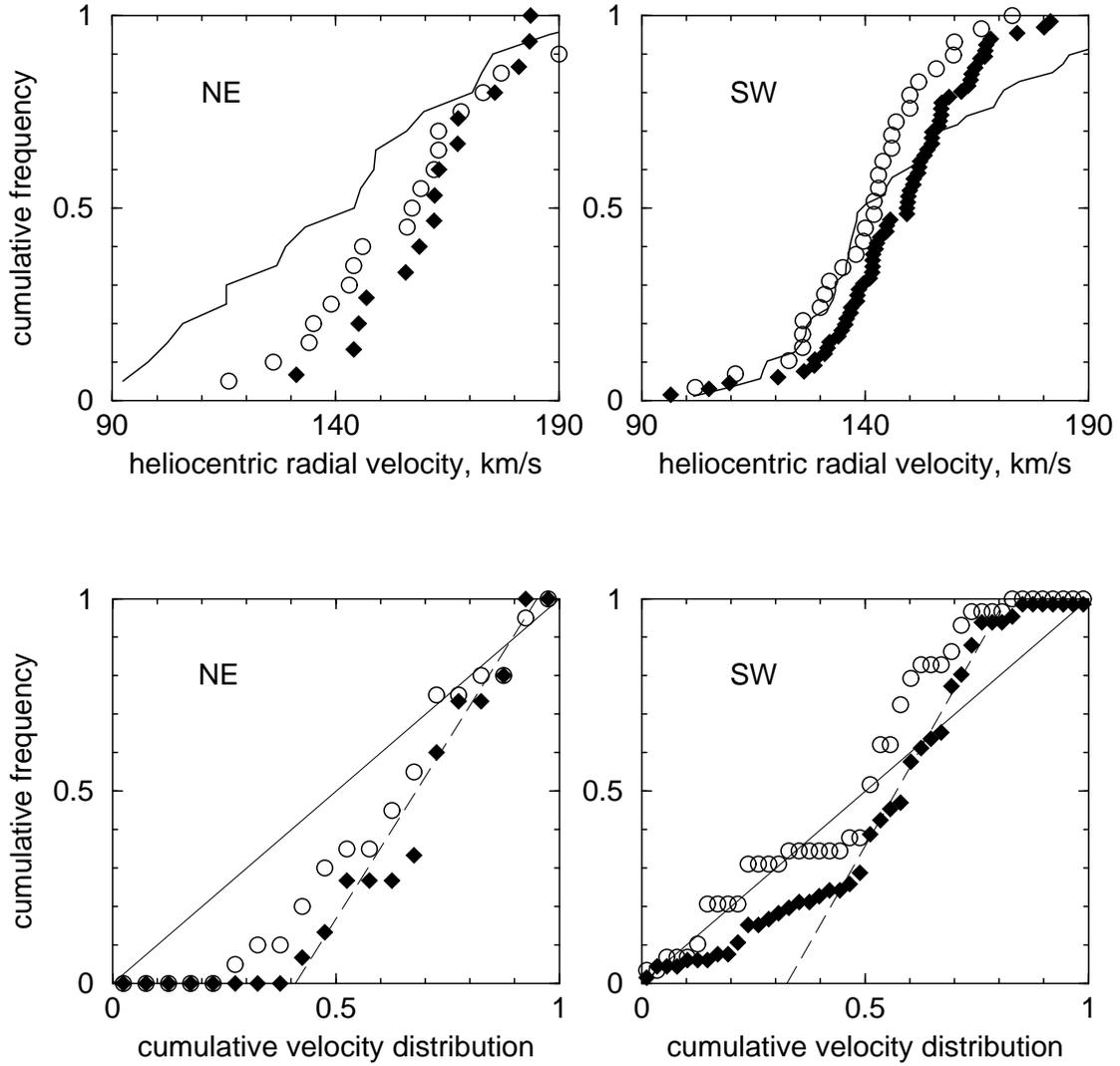,width=0.9\textwidth}}}
\figcaption{
Cumulative distributions of cepheids (open circles), supergiants (filled 
diamonds), and carbon stars (continuous line) toward the NE and SW of the SMC 
bar as a function of the heliocentric radial velocity (upper panels) and the 
carbon star mass distribution (lower panels).  See text. 
\label{fig7}}
\end{figure}

In terms of dynamical processes the relation of the carbon stars to H I gas is 
fundamental, and a comparison to young, recently formed populations introduces
additional uncertainties associated with the idiosyncracies of star formation
during a transient tidal impulse.  Still, (and following the referee's 
recommendation) there is some interest in comparing the distribution with 
young populations that in time formed after the tidal impulse. Figure 7 
provides this comparison, showing the cumulative distribution of cepheids 
(from Mathewson \etal 1988) and supergiants (from Maurice \etal 1989).  The 
upper panels show the cumulative distribution in terms of heliocentric radial 
velocity for the carbon stars, by a countinous line, which differs from that 
of stars formed after the tidal impulse: the cepheids and supergiants, 
indicated by open circles and filled diamonds, respectively.  While the 
asymmetric distribution in the NE bar may result from a unique process, that 
seen in the SW bar represents a puzzle which is not resolved by postulating 
significantly different velocity dispersions for the three populations 
compared. It is worth noting that a dynamical explanation for the observed 
discrepancies, based on differences in intrinsic velocity dispersions, should 
be symmetric about the centers of the plots, at F = 0.5, and such symmetry is 
nowhere in evidence.  The young populations of the NE bar, lying at more 
positive velocities at all positions in the line of sight, appear to occupy the 
rear of the stellar volume mapped by the carbons stars (lower panel). In the 
SW bar no comparably simple explanation for the onset of the distribution 
discrepancy at velocities more positive than 140 km s$^{-1}$ suggests itself. 

\section{\bf SUMMARY AND CONCLUDING REMARKS}
The kinematics of 150 carbon stars on the periphery of the Small Magellanic
Cloud shows an asymmetric distribution characterized by a total absence of
rotation at radii between 3 and 8 kpc.

Beyond 5 kpc the distribution of the carbon stars becomes increasingly
asymmetric, with a marked deficiency in the quadrant centered to the South of
the SMC.  At radii beyond 7 kpc the carbon stars cluster on the position angle
in the direction toward the LMC {PA({\it l,b})=295$^\circ$}.  In the region 
between 10 and 14 kpc the 7 carbon stars found all lie in a 15$^\circ$ sector 
immediately to the South of this same direction, leaving the quadrant to the 
North vacant.

With but two exceptions the velocities of the carbon stars found beyond 5kpc
from the SMC photocenter are seen at the low velocity envelope of the neutral
hydrogen mapped by McGee \& Newton (1986).  Further, we note that the H I
elements of higher density as well as most of the early type stars lie at this
low velocity envelope.

The distribution of material in the ICR is asymmetric about the direction 
toward the LMC in other respects. Beyond 5 kpc the northern ``fan'' (Figure 5, 
upper panel) contains most of the HI and relatively few ICR carbon stars, 
while the southern fan (Figure 5, lower panel) contains most of the ICR carbon 
stars, and less than a quarter of the ICR H I.

In the direction toward the LMC the kinematic asymmetry of material in position
angle is most apparent for PA({\it l,b}) between 270$^\circ$ and 300$^\circ$.  
In this range the slope of the spine of densest H I elements is negative, from 
70 km s$^{-1}$ at PA = 270$^\circ$ to 0 km s$^{-1}$ at PA = 300$^\circ$.  From 
this we infer an inclination of the SMC's orbit about the LMC/SMC barycenter 
of 73$^\circ$ to the plane of the sky.

While the asymmetries at radii beyond 5 kpc from the photocenter arise from a
comparison of carbon star kinematics with those of H I and early type stars, a 
corresponding asymmetry is seen in lines of sight toward the photocenter, in an
 inhomogeneous distribution of the carbon stars observed by Hardy, Suntzeff, \&
Azzopardi (1989) when scaled to the H I distribution mapped by Mathewson, Ford,
\& Visvanathan (1988).  A deficiency of H I is seen on the near side, and a
comparative excess of H I on the far side.  This inhomogeneity is found
incompatible with an homogeneous admixture at a level of 0.03 percent.

The conclusion becomes inescapable that in addition to gravity other forces act
that significantly modify the motion of the gas throughout the entire SMC. One 
is tempted to consider these phenomena in the same context first examined by 
Moore \& Davis (1994), who proposed two mechanisms capable of generating a 
separation akin to that found here. One is the action of a collision impulse 
at pericenter during the LMC/SMC encounter that incrementally decreased the 
orbital momentum of the SMC gas interacting with the LMC disk gas.  That 
interaction should not affect the SMC stars.  The alternative mechanism is the 
continuous action of ram pressure during the motion of the SMC through the 
local hot Galactic corona.  The three body character of the problem defeats 
intuitive insights into deciding which of the two proposed processes 
dominates.  While we may not yet be able to decide the manner in which these 
two mechanisms combine, it seems worth noting that, whatever the combination,
in the end most of the gas of the recent LMC/SMC interaction is found 
preferentially in the tidal tail (while stars dominate in the bridge).  Thus, 
although both the tail and the bridge lie projected on the sky in the ICR, the 
formation events appear to be of the same form as whatever process formed the 
Magellanic Stream proper, and may not require the special binary relation 
between the LMC and SMC postulated as essential so as to deposit debris into 
the ``tail alone'' as proposed by Murai \& Fujimoto (1980).  

Instead, we are reminded, rather, of the syndrome experienced by UGC7636 after 
its interaction with NGC4472 (Sancisi, Thonnard, \& Ekers 1987, and Patterson \& 
Thuan 1993). Here the photographic appearance of UGC7636 shows faint evidence of 
a tail, {\it without} a leading bridge (Arp 1966). The detailed VLA H I map of 
McNamara \etal (1994) shows the H I of UGC7636 confined entirely to the tail 
alone, and photometry of several star clusters that formed in the tail gas after 
the interaction (Lee, Kim, \& Geisler 1997).  Of course, the density of the hot
gas envelope centered on NGC4472 provoking the syndrome seen in UGC7636 is some 
20 times greater than that inferred for the Magellanic Stream (Weiner \& 
Williams 1996, and Sancisi, Thonnard, \& Ekers 1987), and so, quite 
unsurprisingly, the phenomenology seen in the SMC is correspondingly milder.

Nonetheless it seems appropriate to end with a cautionary note that when we
talk of the haloes of dwarf galaxies we refer to dynamically stationary 
settings, where things remain structurally well ordered.  Here our description
of the SMC is of a setting in which recent dynamical violence has disrupted 
whatever order may once have existed.  In such circumstances to talk about a 
halo is likely to mislead our thinking.






\acknowledgments
This project is supported financially, in part (S.D.) by the Natural
Sciences and Engineering Research Council of Canada.



\appendix

\section{Appendicial material}
Star names are based on the field in which they are found, followed by a number
corresponding to our candidate list.  Subsequent columns give: equatorial 
coordinates; the averaged heliocentric radial velocity; the radial velocity
corrected for solar motion ${\it U, V, W}$ = 9, 12, 7 km s$^{-1}$, adopting 
$V_\odot$ = 225km s$^{-1}$; the spectrum quality parameter Q (see Kunkel, 
Demers, \& Irwin (1997) for an explanation); the number of spectra obtained; a 
magnitude; a color; and remarks when appropriate.

\scriptsize
\begin{deluxetable}{clccccrccccc}
\tablecaption{Small Magellanic Cloud periphery carbon stars}
\tablewidth{0pt}
\tablenum{1}
\scriptsize
\tablehead{
\colhead{Name}&\colhead{id field}&\colhead{RA\ \ \ \  (1950)\ \ \ \  Dec}&\colhead{$\ell$}&\colhead{b}&\colhead{$v_{helio}$}&\colhead{$v_{gc}$}&\colhead{Q}&\colhead{n}&\colhead{R}&\colhead{$B_j-R$
}&\colhead{comments}}
\startdata
C0142-7219&029-010& 1 42 03.7  --72 19 48& 318.1&--39.8& 122.0& --9.2 &7& 1 &15.4&3.45\\ 
C0136-7239&029-023& 1 36 16.6  --72 39 55& 298.1&--44.2& 173.3& 15.5 &7& 2 &16.0&3.17\\
C0141-7243&029-031& 1 41 51.3  --72 43 52& 297.5&--44.0& 144.4&--14.6 &7& 1 &15.6&2.53 \\
C0142-7251&029-035& 1 42 48.4  --72 51 48& 297.5&--43.9& 128.9&--30.4 &7& 2 &15.6&4.53\\
C0137-7257&029-039& 1 37 42.9  --72 57 20& 298.0&--43.9& 166.6&  7.9 &7& 1 &15.6&3.41\\
C0134-7301&029-040& 1 34 55.7  --73 01 29& 298.3&--43.9& 174.3& 16.0 &7& 1 &15.8&2.90\\
C0130-7312&029-048& 1 30 32.7  --73 12 55& 298.5&--43.7& 125.9&--32.6 &7& 1 &15.7&2.94\\
C0135-7331&029-066& 1 35 11.3  --73 31 33& 298.5&--43.4& 129.8&--29.4 &7& 1 &15.6&4.10\\
C0144-7347&029-098& 1 44 04.1  --73 47 18& 297.7&--43.0& 125.1&--36.1 &7& 1 &15.4&2.82\\
C0143-7356&029-110& 1 43 59.1  --73 56 03& 297.8&--42.8& 184.7& 23.0 &7& 1 &15.6&2.72\\
C0141-7403&029-123& 1 41 53.1  --74 03 16& 298.0&--42.7& 164.6&  2.9 &7& 1 &15.9&2.65\\
C0134-7505&029-218& 1 34 26.4  --75 05 26& 299.0&--41.9& 131.9&--30.3 &7& 1 &15.6&2.95\\
C0118-7521&029-232& 1 18 04.6  --75 21 11& 300.5&--41.8& 150.5& --9.8 &7& 1 &15.6&2.99\\
C0117-7539&029-243& 1 17 09.5  --75 39 46& 300.6&--41.5& 156.0& --4.8 &7& 1 &16.3&2.93\\
C0149-7525&029-244& 1 49 33.3  --75 25 53& 298.0&--41.3& 137.9&--27.1 &7& 1 &16.8&3.99\\
C0020-7527&029-247& 0 20 39.6  --75 27 13& 305.3&--41.7& 140.8&--12.1 &7& 1 &15.0&3.33\\
C0130-7554&029-259& 1 30 34.5  --75 54 26& 299.6&--41.1& 158.4& --4.8 &7& 1 &15.3&2.61\\
C0142-7549&029-261& 1 42 49.9  --75 49 23& 298.6&--41.0& 125.2&--39.6 &7& 1 &15.8&3.69\\
C0118-7559&029-263& 1 18 12.4  --75 59 57& 300.6&--41.2& 150.9&--10.6 &7& 1 &16.4&4.03\\
C0057-7603&029-265& 0 57 03.5  --76 03 11& 306.8&--40.9& 184.3& 32.2 &7& 1 &16.3&2.57\\
C0055-7603&029-267& 0 55 59.7  --76 03 14& 302.4&--41.2& 169.5& 10.6& 7 &1& 15.8& 2.46\\
C0058-7613&029-268& 0 58 33.2  --76 13 01& 304.2&--41.1& 149.9& --6.2& 7& 1& 15.7& 4.59\\ 
C0138-7609&029-270& 1 38 56.7  --76 09 24& 299.0&--40.8& 151.6&--13.1& 7& 1& 15.6& 2.43\\
C0050-7635&029-278& 0 50 15.6  --76 35 17& 302.8&--40.7& 165.8&  6.5& 7 &1& 15.6& 2.69\\
C0055-7644&029-279& 0 55 17.5  --76 44 35& 302.5&--40.6& 187.4& 27.4& 7 &1& 16.2& 2.44\\
C0036-7650&029-282& 0 36 30.9  --76 50 03& 303.9&--40.4& 138.0&--20.1& 7& 1& 15.4& 2.68\\
C0141-7731&029-289& 1 41 04.3  --77 31 55& 299.3&--39.4& 134.5&--33.0& 5& 1& 15.4& 2.61&  M\\
C0102-7749&029-290& 1 02 15.2  --77 49 38& 302.1&--39.5& 216.0& 60.5& 7 &1 &15.8& 2.72 \\
C0155-7238&030-02 & 1 55 21.0  --72 38 13& 296.2&--43.8&  97.8&--63.3& 7& 1& 15.6& 3.83\\
C0154-7238&030-03 & 1 54 11.7  --72 38 53& 296.3&--43.8& 125.7&--35.3& 7& 1& 14.8& 2.80\\
C0151-7306&030-09 & 1 51 02.1  --73 06 54& 296.8&--43.4& 158.4& --3.0& 7& 1& 14.8& 2.88 \\
C0132-7309&030-11 & 1 32 23.2  --73 09 30& 298.6&--43.8& 172.7& 14.5& 7 &1 &15.5& 3.17 \\
C0155-7343&030-15 & 1 55 29.4  --73 43 16& 296.7&--42.8& 151.2&--11.8& 7& 1& 15.8& 3.31 \\
C0149-7354&030-18 & 1 49 56.2  --73 54 03& 297.3&--42.8& 107.1&--37.5& 7& 1& 15.3& 2.47 \\
C0133-7351&030-23 & 1 33 41.7  --73 51 38& 298.7&--43.1& 122.1&--37.6& 6& 1& 16.3& 2.70\\
C0148-7403&030-24 & 1 48 47.2  --74 03 32& 297.4&--42.6& 153.6& --9.0& 7& 1& 16.3& 3.17 \\
C0135-7356&030-25 & 1 35 08.1  --73 56 40& 298.6&--43.0& 153.6& --6.5& 7& 1& 15.0& 2.83 \\
C0139-7406&030-29 & 1 39 20.0  --74 06 46& 298.3&--42.7& 188.4& 27.1& 4 &1 &16.9& 4.35 & wk C\\
C0147-7437&030-34 & 1 47 18.4  --74 37 33& 297.8&--42.1& 151.1&--12.2& 7& 1& 15.1& 2.80 \\
C0131-7447&030-36 & 1 31 25.2  --74 47 31& 299.2&--42.2& 183.1& 21.8& 5 &1 &14.7& 2.58 & wk C\\
C0127-7452&030-37 & 1 27 49.7  --74 52 03& 299.5&--42.2& 139.1&--21.6& 7& 1& 14.5& 2.94 \\
C0154-7508&030-38 & 1 54 50.3  --75 08 26& 297.4&--41.4& 126.4&--39.1& 7& 1& 15.1& 3.65 \\
C0208-7515&030-40 & 2 08 23.8  --75 15 23& 296.4&--41.0& 134.1&--33.7& 7& 1& 15.1& 2.89\\
C0131-7505&030-42  &1 31 42.3  --75 05 26 &299.3&--42.0 &121.9&--39.6 &7 &1 &15.2 &2.64\\
C0127-7503&030-43  &1 27 26.0  --75 03 02 &299.6&--42.0 &173.7& 12.5 &5 &1 &16.0 &2.50\\
C0201-7536&030-44  &2 01 28.5  --75 36 10 &297.1&--40.9 &130.5&--36.6 &6 &1 &15.0 &2.48&flat sp.\\
C0143-7558&030-47  &1 43 22.0  --75 58 39 &298.6&--40.9 &164.8& --0.3 &7 &1 &14.7 &2.61\\
C0105-6810&051-08  &1 05 53.4 --68 10 34 &300.5&--49.0 &139.6& --2.7 &7 &2 &14.9 &3.16 \\
C0100-6845&051-14  &1 00 20.0 --68 45 21 &301.4&--48.5 &143.2&  0.8 &7 &1 &15.2 &2.43\\
C0055-6847&051-15  &0 55 32.5 --68 47 54 &302.0&--48.5 & 93.8&--47.7 &7 &1 &15.3 &2.76 \\
C0059-6849&051-16  &0 59 53.2 --68 49 23 &301.5&--48.4 & 64.6&--77.9 &7 &1 &15.6 &2.54\\
C0049-6914&051-24  &0 49 21.8 --69 14 29 &302.9&--48.1 &183.3& 41.9 &7 &1 &15.5 &3.31\\
C0122-6910&051-26  &1 22 27.0 --69 10 32 &298.5&--47.8 &164.7& 16.7 &7 &1 &15.2 &2.87\\
C0047-6942&051-31  &0 47 28.5 --69 42 40 &303.1&--47.6 &137.2& --5.1 &7 &1 &15.4 &3.06\\
C0036-6954&051-32  &0 36 33.0 --69 54 12 &304.5&--47.4 &151.2& 10.3 &7 &1 &15.5 &2.56\\
C0115-6954&051-33  &1 15 43.6 --69 54 30 &299.6&--47.2 & 94.6&--53.5 &7 &1 &15.4 &3.55\\
C0116-7003&051-35  &1 16 00.9 --70 03 25 &299.6&--47.1 &156.6&  8.1 &7 &1 &16.0 &2.62\\
C0107-7012&051-40  &1 07 39.0 --70 12 58 &300.6&--47.0 &115.7&--31.6 &7 &1 &15.3 &2.88\\
C0115-7036&051-55  &1 15 49.8 --70 36 23 &299.7&--46.5 &164.2& 14.3 &7 &1 &15.5 &2.50\\
C0124-7105&051-73  &1 24 13.6 --71 05 09 &298.8&--46.0 &168.7& 16.3 &4 &1 &15.2 &2.83& wk C \\
C0124-7106&051-76  &1 24 52.6 --71 06 11 &298.8&--46.0 &115.3&-37.0 &7 &1 &15.3 &2.42\\ 
C0124-6758&052-04  &1 24 26.7 --67 58 33 &298.0&--49.1 &206.2& 61.0 &7 &1 &15.8 &2.79\\ 
C0134-6828&052-06  &1 34 31.3 --68 28 36 &296.7&--48.3 &109.7&-39.0 &7 &3 &15.9 &3.25\\ 
C0112-6825&052-08  &1 12 45.8  --68 25 07 &299.6&--48.7 &157.2& 13.0 &7 &1 &15.9 &2.57 \\
C0117-6848&052-10  &1 17 06.3  --68 48 08 &299.1&--48.3 &119.2&--26.6 &7 &3 &15.0 &3.60 \\
C0121-6923&052-12  &1 21 45.9  --69 23 49 &298.7&--47.6 &136.2&--12.0 &6 &1 &15.9 &4.11 \\
C0110-6932&052-13  &1 10 46.9  --69 32 52 &300.1&--47.6 &132.9&--13.5 &6 &1 &15.9 &3.14 \\
C0127-6842&052-14  &1 27 15.2  --69 42 26 &298.1&--47.2 &209.3& 59.2 &7 &1 &15.8 &2.65 \\
C0132-6956&052-16  &1 32 30.0  --69 56 22 &297.5&--46.9 & 98.8&--52.7 &7 &1 &15.6 &3.97 \\
C0110-6957&052-17  &1 10 02.4  --69 57 05 &300.3&--47.2 &158.0& 10.7 &4 &1 &16.6 &3.12 &faint sp.\\
C0120-7028&052-22  &1 20 18.1  --70 28 10 &299.1&--46.6 &147.4& --2.9 &7 &1 &16.6 &3.44 \\
C0109-7022&052-23  &1 09 41.4  --70 22 02 &300.4&--46.8 &114.6&--33.5 &7 &1 &15.4 &4.49 \\
C0124-7056&052-29  &1 24 32.4  --70 56 00 &298.8&--46.1 &108.4&--43.6 &7 &1 &14.9 &2.71 \\
C0116-7058&052-30  &1 16 15.5  --70 58 53 &298.1&--44.2 &168.2& 10.4 &7 &1 &15.5 &2.77\\
C0125-7121&052-31  &1 25 14.9  --71 21 24 &299.0&--45.7 & 97.5&--55.3 &7 &1 &15.4 &3.08\\
C0123-7137&052-33  &1 23 00.4  --71 37 09 &299.1&--45.4 &109.2&--44.2 &7 &1 &15.2 &2.98\\
C0144-7145&052-35  &1 44 25.3  --71 45 28 &296.9&--44.9 &132.1&--25.4 &7 &1 &14.2 &2.90\\
C0124-7146&052-36  &1 24 45.0  --71 46 17 &299.0&--45.3 &120.9&--32.9 &7 &1 &15.3 &2.70\\
C0117-7143&052-37  &1 17 40.0  --71 43 13 &299.7&--45.4 &102.5&--50.1 &7 &1 &15.1 &3.04\\
C0123-7148&052-38  &1 23 27.5  --71 48 03 &299.1&--45.2 &120.4&--33.5 &7 &1 &15.9 &2.89\\
C0132-7205&052-39  &1 32 39.5  --72 05 06 &298.2&---44.8 &169.8& 13.6 &7 &1 &15.6 &2.97\\
C0131-7211&052-41  &1 31 18.8  --72 11 53 &298.4&--44.7 &118.9&--37.2 &7 &1 &15.3 &2.90\\
C0148-7213&052-42  &1 48 38.2  --72 13 54 &296.6&--44.3 &116.8&--42.6 &7 &1 &15.0 &3.70\\
C0120-7234&052-45  &1 20 03.7  --72 34 33 &299.7&--44.5 &118.0&--36.9 &7 &1 &15.8 &4.15\\
C0154-7238&052-46  &1 54 11.8  --72 38 54 &296.3&--43.8 &127.3&--33.7 &7 &1 &15.2 &2.81\\
C0118-7246&052-49  &1 18 52.9  --72 46 05 &299.8&--44.4 &132.2&--22.8 &7 &1 &15.3 &3.45\\
C0228-7045&053-3   &2 28 51.0  --70 45 25 &291.7&--44.4 &162.8& --1.3 &7 &4 &15.1 &2.94\\
C0108-6743&080-07  &1 08 28.3  --67 43 46 &300.1&--49.5 &130.9&--10.5 &7 &1 &17.1& 3.44 \\ 
RAW-1694&&1 18 39.1  --72 58 46 &300.0&--44.1 &173.1& 17.6 &7 &1 &17.1&\\ 
RAW-1695&&1 18 42.9  --72 41 31 &299.9&--44.4 &149.4& --5.4 &7 &1 &16.9&\\    
RAW-1696&&1 18 52.9  --72 46 06 &299.9&--44.3 &142.9&--12.2 &7 &1 &16.4&\\    
RAW-1697&&1 19 08.1  --73 13 19 &300.0&--43.9 &145.4&--10.5 &7 &1 &17.3&\\   
RAW-1698&&1 19 13.7  --72 52 03 &299.8&--44.2 &182.6& 27.0 &7 &1 &17.7&\\ 
RAW-1699&&1 19 22.6  --73 09 44 &299.9&--44.0 &143.2&--12.6 &7 &1 &17.1&\\ 
RAW-1700&&1 20 21.3  --72 44 43 &299.7&--44.4 &148.1& --7.0 &6 &1 &17.6&\\  
RAW-1701&&1 20 23.1  --72 37 39 &299.7&---44.5 &191.2& 36.2 &7 &1 &17.0&\\ 
RAW-1702&&1 20 24.9  --72 58 10 &299.7&--44.1 &123.2&--32.7 &7 &1 &17.0&\\
RAW-1703&&1 20 29.2  --73 15 24 &299.8&--43.8 &133.5&--23.0 &7 &1 &17.3&\\ 
RAW-1704&&1 20 30.3  --72 58 26 &299.7&--44.1 &162.4&  6.4 &7 &1 &17.3&\\
RAW-1705&&1 20 57.5  --72 57 00 &299.7&--44.1 &119.0&--36.9 &7 &1 &16.3&\\
RAW-1706&&1 21 14.0  --73 01 41 &299.7&--44.1 &160.5&  4.5 &7 &1 &16.3&\\
RAW-1707&&1 21 55.6  --73 21 58 &299.7&--43.7 &124.3&--32.5 &7 &1 &16.5&\\

\enddata
\end{deluxetable}
\normalsize
\clearpage




\clearpage
\end{document}